\documentclass[letterpaper]{jpsj-suppl}
\usepackage{txfonts} 
\usepackage{url}

\title{Light-quark baryon spectroscopy from ANL-Osaka dynamical coupled-channels analysis}

\author{Hiroyuki \textsc{Kamano}}

\inst{Research Center for Nuclear Physics, Osaka University, Ibaraki, Osaka 567-0047, Japan}

\email{kamano@rcnp.osaka-u.ac.jp}

\recdate{}

\abst{
We review our recent efforts for determining resonance parameters associated with 
light-quark baryons ($N^*$, $\Delta^*$, $\Lambda^*$, $\Sigma^*$)
through comprehensive analyses of various meson production reactions off the nucleon
within the ANL-Osaka dynamical coupled-channels approach.
}

\kword{excited nucleons, hyperons, coupled-channels equations, unitarity, resonances, partial-wave analysis, meson production reactions}

\begin{document}
\maketitle

\section{Introduction}

The light-quark baryons, namely, the nonstrangeness $N^*$ and $\Delta^*$ baryons
and the $\Lambda^*$ and $\Sigma^*$ hyperons with strangeness $S=-1$,
provide rich information about QCD in the nonperturbative domain.
A variety of hadron models, such as constituent quark models~\cite{cap} and 
models based on Dyson-Schwinger equations~\cite{anl-dse},
have been proposed to calculate the mass spectrum and form factors of light-quark baryons
and to clarify the role of the confinement and chiral symmetry breaking of QCD
in understanding of the properties of light-quark baryons.
Also, the real energy spectrum of QCD 
under the (anti)periodic boundary condition 
has been computed recently for the light-quark baryon sector 
within the lattice QCD framework (see, e.g., Refs.~\cite{lattice-spec1,lattice-spec2,lattice-spec3}).

A critical nature of excited baryons is that they are unstable against 
the strong interaction and exist only as resonance states in hadron reactions.
Poles of scattering amplitudes in the complex-energy plane
are identified as resonance states,
and the use of a multichannel reaction framework 
is necessary for properly extracting information on 
such resonance states from the reaction data.
In fact, a number of analysis groups including us have performed
comprehensive analyses of the $\pi N$ and $\gamma N$ reaction data
by making use of sophisticated multichannel approaches, such as 
the on-shell $K$-matrix approaches (e.g., Refs.~\cite{bg2012,vrana}) and 
the dynamical-model approaches (e.g., Refs.~\cite{knls13,juelich13,dmt10}),
and they have successfully extracted the parameters 
(complex pole masses and residues, etc.) associated 
with $N^*$ and $\Delta^*$ resonances defined by poles of scattering amplitudes.
Among those studies, the ones with dynamical-model approaches 
have further revealed the crucial role of (multichannel) reaction dynamics 
in understanding the mass spectrum, structure, and dynamical origin of 
baryon resonances (see, e.g., Refs.~\cite{jlss07,sjklms10}).
Similar studies based on a dynamical-model approach have also performed recently
by us for the $\Lambda^*$ and $\Sigma^*$ sector~\cite{kbp1,kbp2}.

In this contribution, we give an overview of our recent efforts on the spectroscopy of 
light-quark baryons, which is based on the so-called ANL-Osaka dynamical coupled-channels 
(DCC) approach.

\section{ANL-Osaka DCC model}

The basic formula of our DCC approach is the coupled-channels integral equations obeyed by 
the partial-wave amplitudes for $a \to b$ reactions~\cite{msl07}
(here we explain our approach by taking the $N^*$ and $\Delta^*$ sector as an example):

\begin{equation}
T^{(J^P I)}_{b,a} (p_b,p_a;W) = 
V^{(J^PI)}_{b,a} (p_b,p_a;W)
+\sum_c \int_C dqq^2 V^{(J^PI)}_{b,c} (p_b,q;W) G_c(q;W) T^{(J^PI)}_{c,a} (p_c,p_a;W).
\label{lseq}
\end{equation}
Here, $p_a$ is the magnitude of the relative momentum for the channel $a$ 
in the center-of-mass frame;
$W$ is the total scattering energy; the superscripts $(J^PI)$ represent the total angular momentum
$J$, parity $P$, and isospin $I$ of the partial wave; and
the subscripts ($a,b,c$) represent the considered reaction channels (the indices associated
with the total spin and orbital angular momentum of the channels are suppressed). 
For the $N^*$ and $\Delta^*$ sector, we have taken into account
the eight channels, $\gamma^{(*)}N$, $\pi N$, $\eta N$, $K\Lambda$, $K\Sigma$, $\pi\Delta$, $\rho N$, and $\sigma N$, where the last three are the quasi two-body channels 
that subsequently decay into the three-body $\pi\pi N$ channel.

\begin{figure}[t]
\centering
\includegraphics[clip,width=0.67\textwidth]{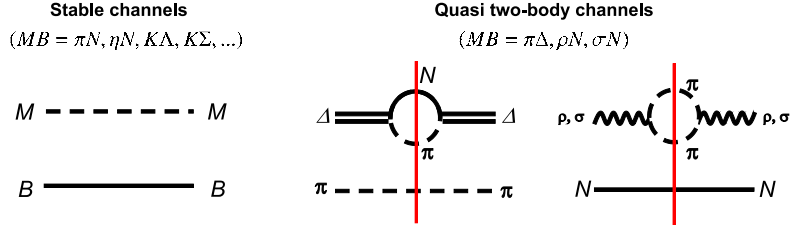}
\caption{\label{green}
Meson-Baryon Green's functions $G_c(q;W)$. 
For the quasi two-body channels, $\pi\Delta$, $\rho N$, and $\sigma N$,
The three-body $\pi \pi N$ cuts are produced in the intermediate processes as 
indicated with the red lines.
The figure is from Ref.~\cite{pos}.
}
\end{figure}
\begin{figure}[t]
\centering
\includegraphics[clip,width=0.85\textwidth]{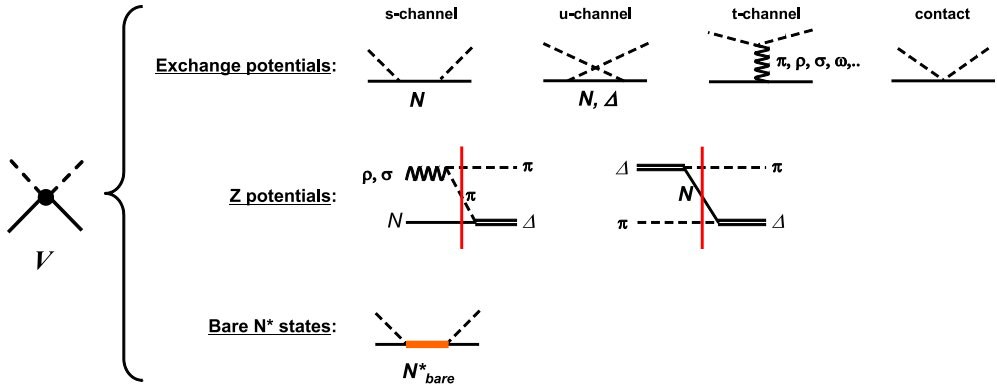}
\caption{\label{potential}
Transition potentials $V_{b,a}^{(J^PI)}(p_b,p_a;W)$.
The three-body $\pi \pi N$ cuts are produced in the $Z$-potentials,
as indicated with the red lines.
The figure is from Ref.~\cite{pos}.
}
\end{figure}

The diagrammatic representation of the Green's functions $G_c(q;W)$ and 
the transition potentials $V^{(J^PI)}_{b,a} (p_b,p_a;W)$ are 
presented in Figs.~\ref{green} and~\ref{potential}, respectively.
Here, the Green's functions for the quasi two-body channels (the right two diagrams 
in Fig.~\ref{green}) and the $Z$-potentials (the middle diagrams in Fig.~\ref{potential})
produce the three-body $\pi \pi N$ cut in the intermediate processes,
and the implementation of both contributions is necessary for maintaining the three-body unitarity.
The s-channel processes mediated by the bare $N^*$ and $\Delta^*$ states 
are also included in our DCC model.
Those bare states couple to the reaction channels through the reaction processes, 
and then become resonance states.
Furthermore, the iterative processes of the exchange potentials 
can also produce resonance poles dynamically.
Our model contains both possibilities in a consistent way.

By solving the coupled-channels integral equation~(\ref{lseq}), 
we can sum up all possible transition processes 
between the considered reaction channels, and this ensures the multichannel two-body 
as well as three-body unitarity for the resulting amplitudes.
Furthermore, off-shell rescattering effects are also taken into account explicitly 
through the momentum integral in Eq.~(\ref{lseq}), which are usually neglected 
in the on-shell approaches.
To extract resonance parameters from the scattering amplitudes given by Eq.~(\ref{lseq}),
one needs to make an analytic continuation of the amplitudes to the (lower half of) 
complex energy plane. 
This can be accomplished by appropriately changing the path of momentum integral $C$ in
Eq.~(\ref{lseq}).
See Refs.~\cite{ssl,ssl2} for the details of the analytic continuation 
method employed for our analysis.

The DCC model for the $\Lambda^*$ and $\Sigma^*$ sector with strangeness $S=-1$
can be constructed in the same way as the $N^*$ and $\Delta^*$ sector by replacing
the reaction channels with $\bar K N$, $\pi \Sigma$, $\pi\Lambda$, $\eta \Lambda$, 
$\pi\Sigma^*$, and $\bar K^* N$, where the last two are the quasi two-body channels
for the three-body $\pi\pi\Sigma$ and $\pi\bar K N$ channels, respectively, 
and by modifying the Green's functions and transition potentials appropriately.

\section{$N^*$ and $\Delta^*$ Spectroscopy through Comprehensive Analysis of 
$\pi N$, $\gamma N$, and $e N$ Reactions}

\begin{figure}[t]
\centering
\includegraphics[clip,width=0.67\textwidth]{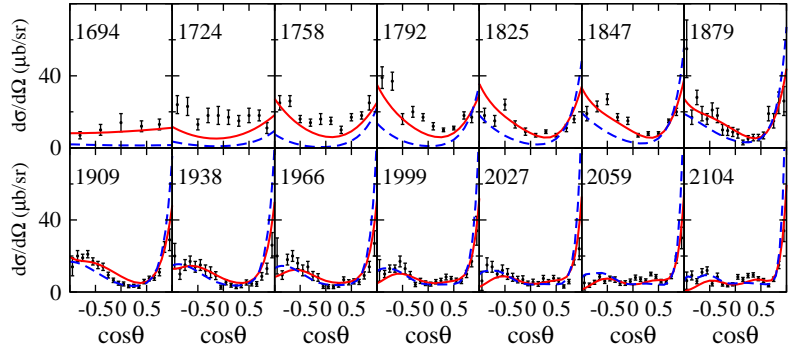}
\caption{\label{pimpk0s0}
Differential cross sections for $\pi^- p \to K^0 \Sigma^0$.
Red solid curves are the preliminary results obtained from the current ongoing analysis,
while blue dashed curves are from the latest published analysis~\cite{knls13}.
The numbers shown in each panel are the corresponding total scattering energy $W$ in MeV.
See Ref.~\cite{knls13} for references of the data.
}
\end{figure}
\begin{figure}[h]
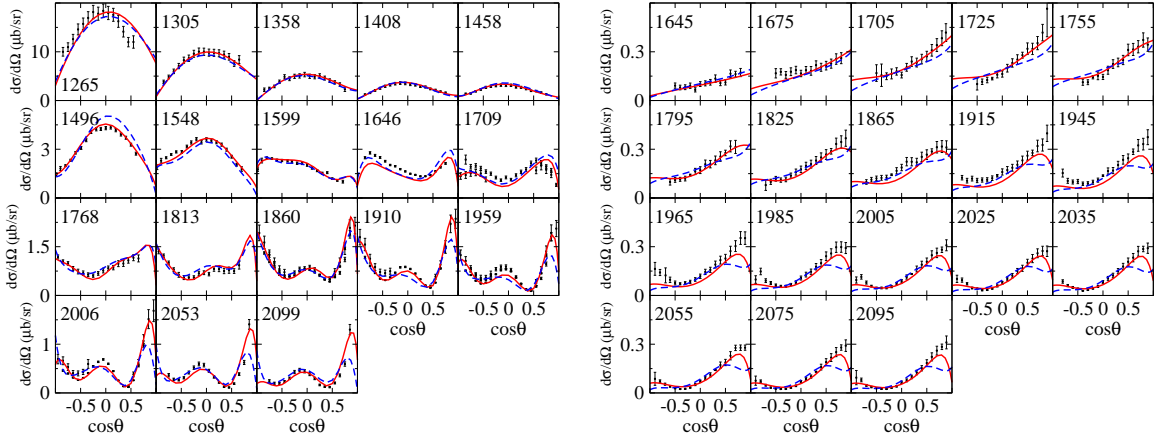

\centering
\includegraphics[clip,width=0.48\textwidth]{gppi0p-dc.eps}
~~~
\includegraphics[clip,width=0.48\textwidth]{gpkpl-dc.eps}
\caption{\label{gp-dc}
(Left) Differential cross sections for $\gamma p \to \pi^0 p$.
(Right) Differential cross sections for $\gamma p \to K^+ \Lambda$.
The numbers shown in each panel are the corresponding total scattering energy $W$ in MeV.
See Ref.~\cite{knls13} for references of the data.
}
\end{figure}

Our latest published model~\cite{knls13} for the $N^*$ and $\Delta^*$ sector
was constructed by performing a comprehensive analysis of unpolarized differential cross sections 
and polarization observables for the $\pi N \to \pi N, \eta N, K \Lambda, K \Sigma$ and 
$\gamma p \to \pi N, \eta N, K \Lambda, K \Sigma$ reactions.
The constructed model covers the energy range from the threshold 
up to $W = 2.3$ GeV for the $\pi N$ scattering and up to $W = 2.1$ GeV for the other reactions.

A couple of results of our fits are presented in Figs.~\ref{pimpk0s0} and~\ref{gp-dc}.
We have been updating our reaction model since the last publication~\cite{knls13}, 
and in the figures the blue dashed curves represent the published results,
while the red solid curves represent the current updated ones.
Some improvements are actually seen in several kinematical regions, 
particularly at low energies of $\pi^- p \to K^0 \Sigma^0$ (Fig.~\ref{pimpk0s0})
and at forward angles of pion and kaon photoproductions (Fig.~\ref{gp-dc}).

\begin{figure}[t]
\centering
\includegraphics[clip,width=\textwidth]{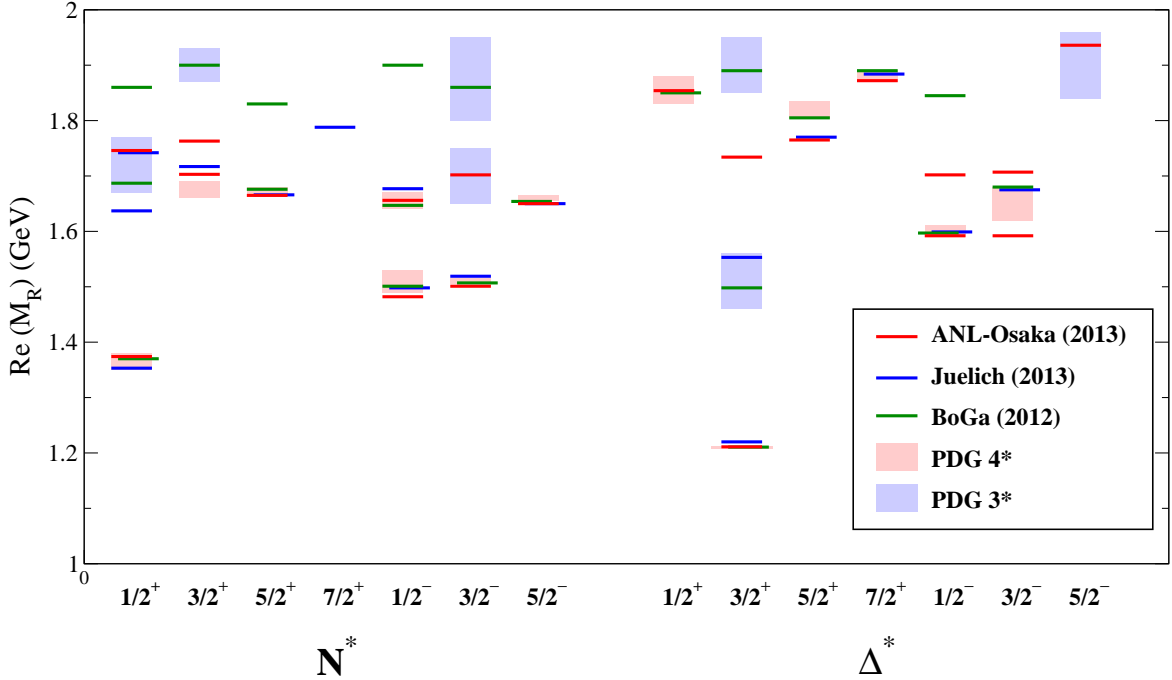}
\caption{\label{nstar-spectrum}
Mass spectra for $N^*$ and $\Delta^*$ resonances.
Real parts of the resonance pole masses $M_R$ are plotted.
Also, only the resonances with $0 < -{\rm Im}(M_R) < 0.2$ GeV are presented.
The results are from (Red) ANL-Osaka (ours)~\cite{knls13}, 
(Blue) J\"ulich~\cite{juelich13}, and (Green) Bonn-Gatchina~\cite{bg2012}.
The spectrum of four- and three-star resonances rated by PDG~\cite{pdg2014} 
is also presented with the red and blue filled squares, respectively, 
of which the length in the longitudinal direction represents 
the range of the real parts of the resonance pole masses assigned by PDG.
}
\end{figure}
In Fig.~\ref{nstar-spectrum}, the mass spectra for the $N^*$ and $\Delta^*$ 
resonances extracted by multichannel analysis groups are presented.
Our spectrum~\cite{knls13} shown in red are compared with 
the ones extracted by the J\"ulich group in 2013 (blue)~\cite{juelich13}
and the Bonn-Gatchina group in 2012 (green)~\cite{bg2012},
and also with the four- and three-star resonances assigned by PDG~\cite{pdg2014}.
The results show that the existence and mass values of low-lying resonances 
have been well determined for most of the spin-parity states.
Thus establishing the spectrum of high-mass resonances
will be a next important task in the $N^*$ and $\Delta^*$ spectroscopy. 
Here it is noted that the J\"ulich group has updated their mass spectrum recently, and it can be found in Ref.~\cite{debora}.

The high-mass $N^*$ and $\Delta^*$ resonances are expected to couple strongly to the three-body $\pi\pi N$ channel.
This can be seen from the partial decay widths evaluated within an earlier version of our 8-channel DCC analysis 
(see Fig.~6 of Ref.~\cite{e45-proposal}),
which actually shows that the high-mass resonances decay dominantly to the $\pi\pi N$ channel.
It is worth mentioning that the second $P_{33}$ resonance, $\Delta(1600)3/2^+$ 
in the notation of PDG~\cite{pdg2014}, also has a large partial decay width 
to the $\pi\pi N$ channel. 
The double-pion production data are therefore 
a key essential to establishing high-mass resonances 
as well as the Roper-like state of the $\Delta$ baryon.
However, so far essentially no differential cross section data 
that can be used for the detailed partial-wave analysis 
were available for the $\pi N \to \pi \pi N$ reactions at high energies, 
and this was a problem for the $N^*$ and $\Delta^*$ spectroscopy.
But now the situation is being improved by HADES~\cite{hades} and J-PARC~\cite{e45}.
In particular, with the J-PARC E45 experiment~\cite{e45-proposal}, it is expected that 
the world data of the $\pi N\to \pi\pi N$ reactions is increased by a factor of 100 or more.

Another important task in the $N^*$ and $\Delta^*$ spectroscopy is 
to determine electromagnetic transition form factors 
between the nucleon and the $N^*$ or $\Delta^*$ resonance.
By studying the form factors, we could see how the transition 
between the effective degrees of freedom describing baryons 
occurs with changes in $Q^2$.
To determine the $Q^2$ dependence of the form factors, one needs 
to analyze meson electroproduction reactions.
So far, we have made such analyses for the data for $p(e,e'\pi)N$
up to $Q^2 = 1.5$ GeV$^2$ within our previous 6-channel DCC model~\cite{jklmss09,ssl2}, 
and, more recently, up to $Q^2 = 3$ GeV$^2$ within our latest 8-channel DCC model~\cite{nks15}.

\begin{figure}[t]
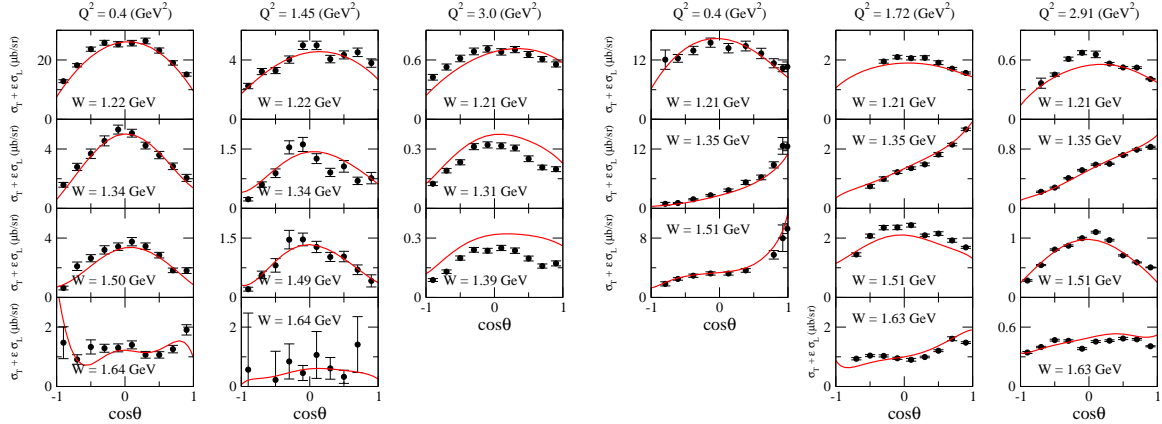

\centering
\includegraphics[clip,width=0.48\textwidth]{epp0p.eps}
\ \ \ \ 
\includegraphics[clip,width=0.48\textwidth]{epp-n.eps}
\caption{\label{elptel}
Structure functions $\sigma_T + \epsilon \sigma_L$ at several $Q^2$ and $W$ values..
The left (right) three columns are for $ep \to e'\pi^0p$ ($ep \to e'\pi^+n$).
The structure function data are from Refs.~\cite{sj,joo,ungaro}.
}
\end{figure}

In Fig.~\ref{elptel}, a couple of results of our recent analysis for the $p(e,e'\pi)N$ data
performed in Ref.~\cite{nks15} are presented.
Here we have used the structure functions as the data to analyze~\cite{sj,joo,ungaro}, 
rather than the original five-fold differential cross sections.
The results capture an overall shape of the structure functions data, 
but it is still not sufficient for the purpose of $N^*$ form factor studies.
Here it is noted that the analysis in Ref.~\cite{nks15} is dedicated for studying the neutrino 
reactions, and therefore our model parameters are not fine-tuned 
for the purpose of studying $N^*$ and $\Delta^*$ transition form factors.
More elaborated analysis of single pion electroproductions is ongoing for the $Q^2$ 
region up to 6 GeV$^2$, and the results will be presented elsewhere.

\begin{figure}[t]
\centering
\includegraphics[clip,width=0.35\textwidth]{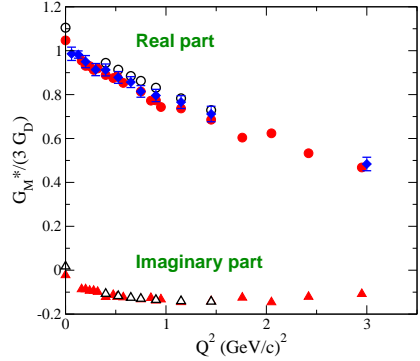}
\caption{\label{gm}
The $Q^2$ dependence of the $M1$ form factor, $G^*_M(Q^2)$, for 
the $\gamma^* p \to \Delta(1232)3/2^+$ transition 
evaluated at the pole position of $\Delta(1232)3/2^+$.
Filled (open) circles are the real parts of $G^*_M(Q^2)$ obtained from 
our 8-channel~\cite{nks15} (6-channel~\cite{jklmss09,ssl2}) model,
while filled (open) triangles are the corresponding imaginary parts.
The diamonds are extracted by experimental groups using the Breit-Wigner 
parametrization~\cite{exp-gm}.
$G_D$ is a dipole factor, $G_D = [1+(Q^2/\Lambda^2)]^{-2}$  with $\Lambda = 0.71$~GeV.
}
\end{figure}

In Fig.~\ref{gm}, the extracted $M1$ transition form factors
between the nucleon and $\Delta(1232)3/2^+$ resonance are presented.
Here it is noted that in our analyses the transition form factors are 
evaluated at the pole positions of the resonances, and thus they inevitably become 
complex because of the fact that resonances are decaying particles.
This is in contrast to the form factors extracted by experiment groups, 
where the phenomenological Breit-Wigner parametrizations are used 
and the extracted values are real.
We find that for the $\Delta(1232)3/2^+$ case, the imaginary parts of the form factors 
are small and the Breit-Wigner results seem close to the real parts of the form factors 
defined by poles.
However, for the higher resonances, the imaginary parts can be comparable 
with the real parts, and in such cases the correspondence between the form factors 
defined by poles and by the Breit-Wigner parametrizations becomes unclear.
The clarification of those differences requires further investigations.

\begin{figure}[t]
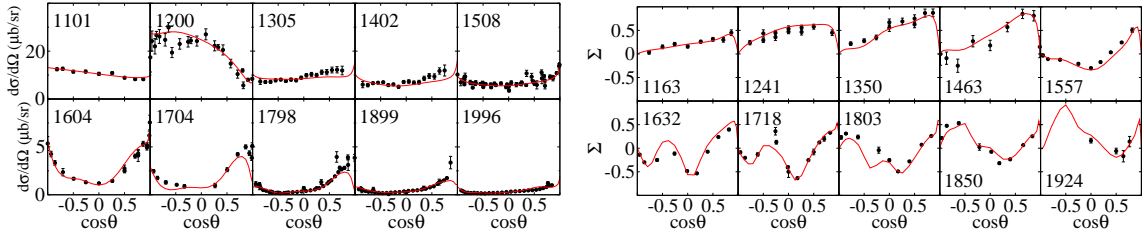

\centering
\includegraphics[clip,width=0.48\textwidth]{gnpi-p_dc.eps}
\ \
\includegraphics[clip,width=0.48\textwidth]{gnpi-p_s.eps}
\caption{\label{gn-obs}
Differential cross sections (left) and photon asymmetries (right)
for $\gamma n \to \pi^- p$.
The numbers shown in each panel are the corresponding total scattering energy $W$ in MeV.
The data are taken from Ref.~\cite{gwu}.
}
\end{figure}

Recently, we have also made an analysis of the data for the single pion 
photoproduction off the ``neutron'' target (Fig.~\ref{gn-obs}).
Analysis of both proton- and ``neutron''-target photoproductions is 
necessary for decomposing the electromagnetic currents into the isoscalar and isovector currents
and determining the electromagnetic interactions of the $N^*$ resonances that have isospin $1/2$.
It is noted that such isospin currents are also 
necessary for studying neutrino-induced reactions (see e.g., Refs.~\cite{nks15,nakamura}).
Currently we use the ``neutron''-target data extracted by other analysis groups from 
the deuteron-target reactions. 
However, in the future we need to analyze the deuteron-reaction data directly and 
extract the $\gamma n \to N^*$ helicity amplitudes in a fully consistent way in our approach.

\section{$\Lambda^*$ and $\Sigma^*$ Spectroscopy through 
Comprehensive Analysis of $K^-p$ Reactions}

The $Y^*$ ($= \Lambda^*, \Sigma^*$) resonances are much less understood than the $N^*$ and $\Delta^*$ resonances.
This can be seen, for example, from the fact that for the $Y^*$ resonances 
only the so-called Breit-Wigner masses and widths had been listed by PDG before 2012~\cite{pdg2012}.
This was a rather unsatisfactory situation~\cite{footnote1} because the Breit-Wigner parameters are nothing more than
``approximation'' of the resonance parameters defined by poles of scattering amplitudes 
in the complex energy plane~\cite{footnote2},
where the latter has a clear physical meaning: the resonance states defined by poles are associated with 
the exact (complex) energy eigenstates of the {\it full} Hamiltonian of the system under the purely outgoing 
boundary condition (see, e.g., Refs.~\cite{madrid1,madrid2}).

In this situation, we have recently made a comprehensive partial-wave analysis of 
the available $K^- p$ reaction data within our DCC approach~\cite{kbp1,kbp2}.
This was accomplished by developing a DCC model for strangeness $S=-1$ sector, which takes into account 
couplings between the two-body $\bar K N$, $\pi\Sigma$, $\pi\Lambda$, $\eta\Lambda$, and $K\Xi$ channels and 
the three-body $\pi \pi \Lambda$ and $\pi \bar K N$ channels that have resonant components of 
$\pi \Sigma^*$ and $\bar K^* N$, respectively.
The model parameters are then determined by fitting to all available data of 
$K^- p \to \bar K N,\pi\Sigma,\pi\Lambda,\eta\Lambda,K\Xi$ reactions from the threshold up to $W=2.1$ GeV.
Our analysis includes the data of both unpolarized and polarized observables, and this results in fitting more than 17,000 data points.
From this analysis, we have successfully determined the partial-wave amplitudes not only for $S$ wave but also $P$, $D$ and $F$ waves,
and also extracted the $Y^*$ mass spectrum defined by poles of scattering amplitudes.
The full details of the analysis and the extracted $Y^*$ resonance parameters can be found in Refs.~\cite{kbp1,kbp2}, and in the following
we will present a highlight of them.

\begin{figure}[t]
\centering
\includegraphics[clip,width=0.75\textwidth]{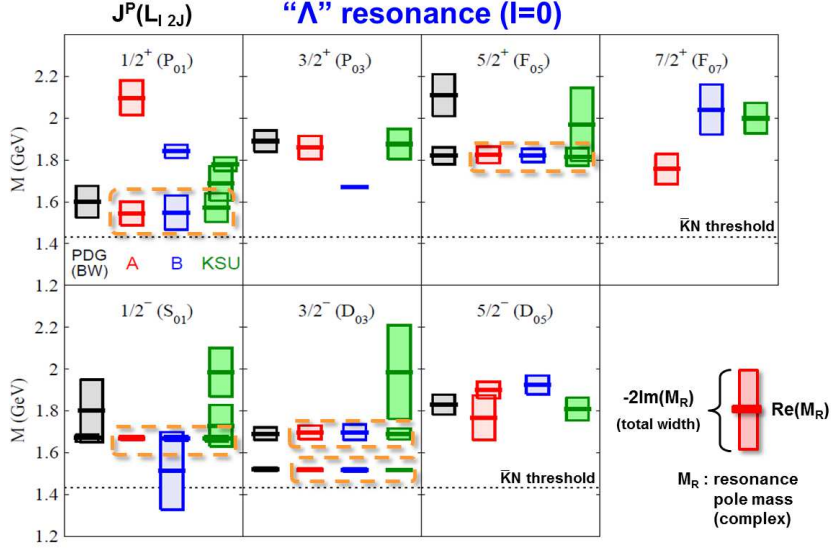}
\caption{\label{lmass}
Extracted mass spectra of $\Lambda^*$ resonances.
Here only the resonances of which complex pole mass has a value satisfying
$m_{\bar K} + m_N \leq {\rm Re}(M_R) \leq 2.1$ GeV and $0\leq-{\rm Im}(M_R) \leq 0.2$ GeV, are presented [$m_{\bar K}$ ($m_N$) is the antikaon (nucleon) mass].
The mass spectra extracted from our two analyses, 
Model~A (red) and Model~B (blue) constructed in Ref.~\cite{kbp1}, 
are compared with the one from the KSU analysis~\cite{zhang2013} (green). 
The Breit-Wigner masses and widths of the four- and three-star resonances rated by PDG~\cite{pdg2014} (black) are also presented.
The well-determined resonances are enclosed with the orange dashed circles.
}
\end{figure}

In Fig.~\ref{lmass}, we compare the mass spectra of $\Lambda^*$ resonances extracted from our analysis~\cite{kbp1,kbp2} 
and the analysis by the Kent State University (KSU) group~\cite{zhang2013}.
In our analysis, we found two distinct solutions that have quite different values for our model parameters, yet both give similar quality of 
the fits to the $K^-p$ reaction data included in our analysis. 
We call them Model A and Model B, and 
their resulting mass spectra are presented in red and blue, respectively.
In the same figure, the spectrum of four- and three-star 
resonances assigned by PDG is also presented.
However, it is noted that the mass spectra of our two models 
and the KSU analysis are the ones given as poles of 
scattering amplitudes, while the PDG values are of the Breit-Wigner masses and widths.
We see that the spectra extracted from our two models and the KSU analysis 
show an excellent agreement for several resonances, 
but in overall, they are still fluctuating between the three analyses.
For example, a $J^P=3/2^+$ $\Lambda$ resonance with a mass 
${\rm Re}(M_R)\sim 1.86$ GeV
is found in Model A and the KSU analysis, while not in Model B
(panel for $J^P=3/2^+$ spectra of Fig.~\ref{lmass}).
If this resonance corresponds to the four-star $\Lambda(1890)3/2^+$ of PDG, 
then this may be one example showing that a four-star resonance rated by PDG 
using the Breit-Wigner parameters is not confirmed by the analyses 
in which the resonance parameters are extracted at pole positions.
As already discussed and emphasized in Refs.~\cite{kbp1,kbp2}, 
this kind of analysis dependence would originate
from the fact that the existing $K^-p$ reaction data are not sufficient 
to eliminate such dependence on the extracted mass spectrum.

\begin{figure}[t]
\centering
\includegraphics[clip,width=0.75\textwidth]{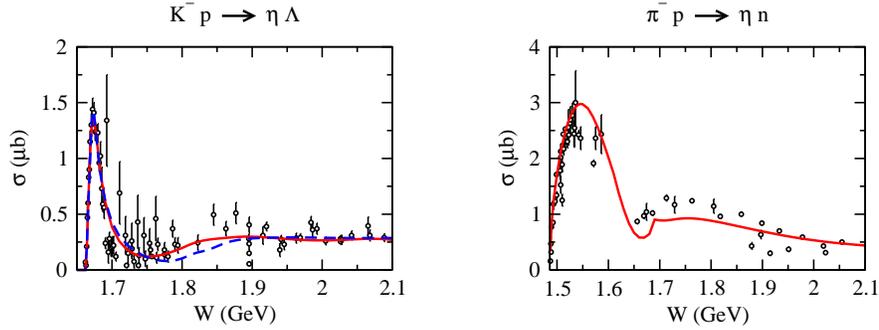}
\caption{\label{eta-tcs}
(Left) Total cross section for $K^- p \to \eta \Lambda$. 
Red solid and blue dashed curves are from Model A and Model B~\cite{kbp1}, respectively.
(Right) Total cross section for $\pi^- p \to \eta n$, where the curve is from our published DCC model for $\pi N$ and $\gamma N$ reactions~\cite{knls13}.
}
\end{figure}

We can see from the lower left-most panel of Fig.~\ref{lmass} 
(spectra for $J^P=1/2^-$ $\Lambda$ resonances) that 
all of the three analyses find a narrow $J^P=1/2^-$ $\Lambda$ resonance located close to 
the $\eta \Lambda$ threshold, $W \sim 1.67$ GeV: $M_R = 1669^{+3}_{-8} -i (9^{+9}_{-1})$ MeV 
for Model A, $M_R = 1667^{+1}_{-2} -i (12^{+3}_{-1})$ MeV for Model B, and 
$M_R = 1667 -i 13$ MeV for the KSU analysis.
This resonance is known as $\Lambda(1670)1/2^-$ and found to be responsible 
for the sharp peak in the $K^- p \to \eta \Lambda$ total cross section near the threshold (left panel of Fig.~\ref{eta-tcs}). 
This behavior looks similar to $N(1535)1/2^-$ in 
the $\pi N \to \eta N$ reaction (right panel of Fig.~\ref{eta-tcs}),
where the contribution from $N(1535)1/2^-$ dominates the peak of the $\pi N \to \eta N$ total cross section near the threshold.

\begin{figure}[t]
\centering
\includegraphics[clip,width=0.65\textwidth]{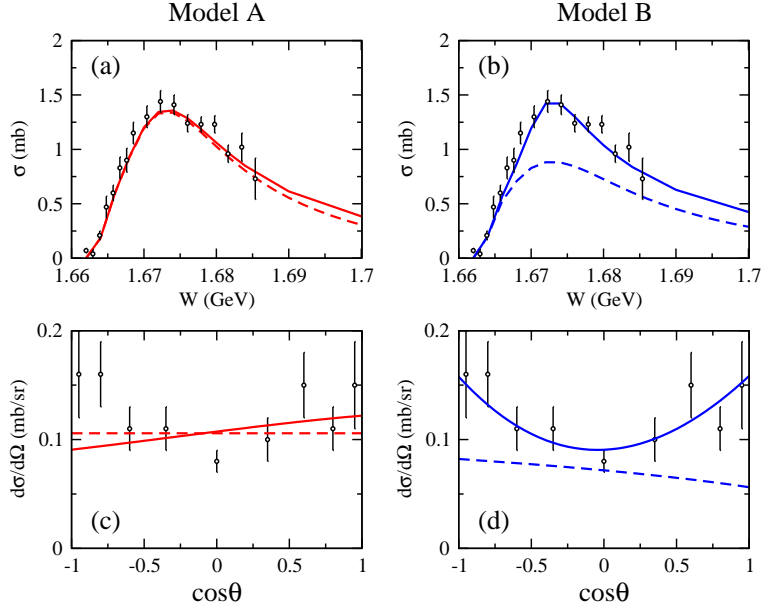}
\caption{\label{kmpetl-tcsdcs}
Total cross section near the threshold (upper panels) and differential cross section at $W = 1672$ MeV (lower panels)
for $K^- p \to \eta \Lambda$
Left panels (right panels) are the results from Model A (Model B). 
Solid curves are the full results, while the dashed curves are the results for which 
the contribution from the $P_{03}$ partial wave is turned off.
The dashed curves are almost dominated by the $S_{01}$ partial wave that contains $\Lambda(1670)1/2^-$.
For Model B, the difference between the solid and dashed curves is due to 
the new narrow $P$-wave $J^P = 3/2^+$ $\Lambda$ resonance.
}
\end{figure}

It is also interesting to see that Model B has another very narrow resonance 
with $J^P =3/2^+$ and $M_R = 1671^{+2}_{-8}-i(5^{+11}_{-2})$ MeV 
(see the panel for $J^P=3/2^+$ spectra in Fig.~\ref{lmass}), 
which has almost the same ${\rm Re}(M_R)$ value as $\Lambda(1670)1/2^-$.
However, currently this resonance is found only in Model B.
Actually, in Model A the peak of the $K^- p \to \eta \Lambda$ total cross section near the threshold 
is completely dominated by $\Lambda(1670)1/2^-$ [Fig.~\ref{kmpetl-tcsdcs}(a)], while in Model B, about 40\% of the magnitude 
of the peak is turned out to come from this narrow $P$-wave $J^P=3/2^+$ $\Lambda$ resonance [Fig.~\ref{kmpetl-tcsdcs}(b)].
Since both models reproduce the total cross section well, it is hard to judge whether 
this new $P$-wave $\Lambda$ resonance should exist or not, as far as looking at the total cross section only. 
However, we can get a deeper insight by looking at differential cross sections.
The lower panels of Fig.~\ref{kmpetl-tcsdcs} show the differential cross section of $K^- p \to \eta \Lambda$ at 1672 MeV,
which corresponds to the peak energy of the total cross section near the threshold.
We see that the differential cross section data show a clear concave-up angular dependence, 
which cannot be described by the $S$-wave amplitudes. 
In fact, we find that Model A, for which the total cross section is dominated by the $S$ wave, 
does not reproduce the angular dependence well.
On the other hand, in Model B, the new $P$-wave $J^P=3/2^+$ $\Lambda$ resonance is responsible for the reproduction of the data, 
suggesting that this angular dependence of the data seems to favor this new resonance.

\section{Summary and Prospects}

We have performed comprehensive partial-wave analysis for the data of various meson production reactions off the nucleon 
within the ANL-Osaka DCC approach. 
We then have successfully extracted the resonance parameters associated with 
the light-quark baryons ($N^*$, $\Delta^*$, $\Lambda^*$, $\Sigma^*$), which are defined 
by poles of scattering amplitudes in the complex energy plane.

We may say that a recent progress on the light-quark baryon spectroscopy 
triggered by multichannel analysis groups is quite remarkable. 
However, a visible analysis dependence still exists in the extracted resonance parameters.
To eliminate such dependence, one would need not only to make further improvements of the analysis methods
of each analysis group, but also to have more extensive and accurate data of meson production reactions 
including the polarization observables (see, e.g., Ref.~\cite{shkl11}).
Regarding this, there were a lot of contributions on the experimental activities at
the electron, photon, and hadron beam facilities to this NSTAR2015 workshop, and 
a variety of new or planned experiments were reported.
With the help of these experiments, we would be able to make further
progress towards understanding nonperturbative nature of the low energy QCD.

Finally, the framework of our DCC approach itself is quite general, and it has been applied not only to
the light-quark baryon spectroscopy, but also to the neutrino-induced reactions~\cite{knls12,nks15} associated with 
the neutrino-oscillation experiments in the multi-GeV region and the meson spectroscopy~\cite{knls11,knls12-2}.
We plan to put more efforts into these directions, too.
\\

The author would like to thank T.-S.~H.~Lee, S.~X.~Nakamura, and T.~Sato for their collaborations.
This work was supported by the Japan Society for the Promotion of Science (JSPS) KAKENHI Grant No. 25800149.
The author also acknowledges the support of the HPCI Strategic Program (Field~5
``The Origin of Matter and the Universe'') of Ministry of Education, Culture, Sports, Science
and Technology (MEXT) of Japan.


\begin{thebibliography}{99}
\bibitem{footnote1}
This is in contrast to the $N^*$ and $\Delta^*$ cases,
for which the resonance parameters defined by poles have also been 
extensively studied by a number of multichnal analysis groups, and 
both the pole and Breit-Wigner results have been listed by PDG for three decades.
\bibitem{footnote2}
This ``approximation'' often does not work well for the resonances that have a large inteference with backgrounds and/or 
are not well isolated from the other resonance poles and the singularities (branch points etc.)
in the complex energy plane.
For example, the mass of the Roper resonance is often referred to as $\sim$1440 MeV. 
However, it is the value obtained within the Breit-Wigner parametrization. 
In the pole definition, (the real part of) the mass of the Roper resonance is $\sim$1370 MeV, i.e., $\sim 70$ MeV lower than
the Breit-Wigner mass.
\end{thebibliography}
\end{document}